\documentstyle[sprocl,epsf,psfig]{article}

\def\be{\begin{equation}}
\def\ee{\end{equation}}
\def\bea{\begin{eqnarray}}
\def\eea{\end{eqnarray}}
\newcommand \beq{\begin{eqnarray}}
\newcommand \eeq{\end{eqnarray}}
\newcommand{\set}[2]{\newcommand{#1}{#2}}
\set{\pa}{\partial \over \partial\, }
\set{\leftvector}{\stackrel{\leftarrow}{\partial }}
\set{\rightvector}{\stackrel{\rightarrow}{\partial }}

\begin{document}

\title{Formation of correlations at short time scales and consequences on interferometry methods}

\author{K. Morawetz}
\address{
Fachbereich Physik, Universit\"at Rostock,
18051 Rostock, Germany}
\author{H. S. K\"ohler}
\address{Physics Department,
University of Arizona,Tucson,Arizona 85721}


\maketitle\abstracts{The formation of correlations due to collisions in an
interacting nucleonic system is investigated shortly after a disturbance.
Results from one-time kinetic equations are compared with the
Kadanoff and Baym two-time equation with collisions included in
second order Born approximation.
A reasonable agreement is found for a proposed
approximation of the memory effects by a finite duration of
collisions. The formation of correlations and the build up time is calculated analytically for the high
temperature and the low temperature limit. This translates into a
time dependent increase of the effective temperature on time
scales which interfere with standard fire ball scenarios of heavy
ion collisions. The consequences of the formation of correlations on the two- particle interferometry are
investigated and it is found that standard extracted lifetimes should be corrected
downwards.}

\section{Introduction}

To extract information about the space- time character of
particle- emitting sources is one of currently much investigated
tasks \cite{BDH94,BGJ90}. The size of stars are found to be measurable
by coincident
photons \cite{BT56}. In laboratory this method is used to analyse
heavy ion reactions \cite{P87}. The two-particle correlation function
carries the information about nuclear interaction and correlations in the
emitting source \cite{Ko77,G91} and is given by the ratio of two -
and single particle emission probability
\beq
R(P=p_1+p_2,q=(p_1-p_2)/2)+1={\Pi(p_1,p_2) \over \Pi(p_1)\Pi(p_2)}
\eeq
which can alternatively be expressed by cross sections
\cite{B87}. This two-particle correlation function can be
written in terms of the the two-particle wave function
$\varphi(q,r)$
\beq
R(P,q)+1=\int d^3 r F_p(r) |\varphi(q,r)|^2
\eeq
while the average of the single particle Wigner functions of the
source is given by \cite{PT87,BGPG91}
\beq
F_p(r)={\int d^3 R \,  f(P/2,R+r/2,t^>) \, f(P/2,R-r/2,t^>) \over \left (\int
d^3R \,f(P/2,R,t^>)\right )^2}.\label{fp}
\eeq
Here $f(p,r,t>)=\int\limits^{t^>} dt \, g(p,r-p(t^>-t)/m,t)$ is the phase-space
Wigner distribution of particles
of momenta $p$ at position $r$ at some time $t^>$ after emission with probability $g$.
A rigorous derivation of this formula with the discussion of the necessary
neglects has been given in \cite{BDH94} for the case of initially
uncorrelated particles $|\varphi(p,r)|^2\propto 1\pm \cos(rq)$ corresponding to 
bosons/fermions.
Eq. (\ref{fp}) is well suited for simulation of heavy ion
reactions where the one-time distribution
function $f({\bf r,p},t)$ is then determined by solving
appropriate kinetic equations (BUU) or simulating equation of motions (QMD).
While most current codes rely on the quasiclassical Boltzmann
equation including Pauliblocking effects, a quantum two-time theory for the
time-evolution of real time Green's functions $G({\bf r,p},t,t')$ has been
developed using the Schwinger-Keldysh formalism already 30 years
ago. The quantum image of the
classical Boltzmann equation is usually referred to as the Kadanoff-Baym
(KB)
equations \cite{KB62}. These equations have often been considered too
complicated to solve numerically in the past. However, several numerical applications exist now
\cite{hsk96}.

These kinetic equations describe different relaxation stages.
During the very fast first stage, correlations imposed by the initial
preparation of the system
are decaying \cite{B46,BKSBKK96}. These are contained in
off-shell or dephasing processes described by two-time propagators. During this stage of
relaxation the quasiparticle picture is established \cite{LKKW91,MSL97a}.
After this very fast process the second state develops
during which the one-particle distribution relaxes towards the equilibrium
value with a relaxation time.
This is characterized by a nonlocal kinetic equation in agreement
with the virial correction to the approached equation of state
\cite{SLM96}.
We will focus on the first stage
which is related to the formation of correlations and will find a
measurable effect on the two-particle correlation function.
The extracted lifetimes are shown to be too high if one
ignores this transient time effects. This can be considered to
belong to the initial state correlations. While in
\cite{A92,ARS93} the initial state correlations are assumed to be
of equilibrium type and discussed by density and temperature
dependence we like to investigate here a nonequilibrium effect
which consists in the fact that correlations need a specific time
to be build up.

The formation of correlations is connected with an increase of the kinetic energy or equivalently the build
up of correlation energy. This is due to rearrangement processes which let decay higher order correlation
functions until only the one - particle distribution function relaxes.

\section{Kinetic description}

The time dependence of the kinetic energy as a one-particle observable will
be investigated within the kinetic theory. This can only be accomplished if we employ a kinetic equation which
leads to total energy conservation. It is immediately obvious that the ordinary Boltzmann
equation cannot be appropriate for this purpose because the kinetic energy is in this case an invariant of the collision
integral and constant in time. In contrast, we have to consider
non-Markovian kinetic equations of Levinson type \cite{L65,L69}, which account for the formation of
two particle correlations and which conserves the total energy \cite{M94}
\cite{JW84,MWR93,HJ96}
\begin{eqnarray}
&&\frac{\partial}{\partial t}f_a(t)=\frac{2 s_a}{\hbar^2}\sum\limits_b s_b
\int\frac{dpdq}{(2\pi\hbar)^6}V_{\rm D}^2(q)\nonumber\\
&&\times
\int\limits_{t_0}^t d\bar t\,
\exp\left\{-{t-\bar t\over\tau_g}\right\}\,
{\rm cos}\left\{\frac{1}{\hbar}(t-\bar t)\Delta_E\right\}
\left\{f'_a f'_b \bar f_a \bar f_b-
f_a f_b \bar f'_a \bar f'_b\right\},
\label{kinetic}
\end{eqnarray}
where $\Delta_E={k^2\over 2m_a}+{p^2\over 2m_b}-{(k-q)^2\over 2m_a}-
{(p+q)^2\over 2m_b}$ denotes the energy difference between initial and final
states. The retardation of distributions, $f_a(k,\bar t)$,
$f'_a(k-q,\bar t)$ etc., is balanced by the quasiparticle damping $\tau_g$
and $\bar f = 1-f$, the free particle dispersion
$E=p^2/2m$ and the spin-isospin degeneracy $s$.
The distribution functions are normalized to the density as $\int {d p\over (2 \pi \hbar)^3} f(p)=n/s$.

The Boltzmann collision integral is obtained from
equation (\ref{kinetic})
if: (i) One neglects the time
retardation in
the distribution functions, i.e. the memory effects. This would lead to
gradient contributions to the kinetic equation which can be shown to be
responsible for the formation of high energy tails in the distribution
function \cite{MR95,SL95} and virial corrections \cite{SLMa96,SLMb96,SLM96}.
This effect will be established on the second stage of
relaxation.
(ii) The finite initial time $t_0$ is set equal to
$-\infty$ corresponding to what is usually referred to as
the limit of complete
collisions.
This energy broadening or off-shell behavior in (\ref{kinetic}) is
exclusively related to the spectral properties of the one-particle
propagator and therefore determined by the relaxation of two-particle
correlation. 

Since we are studying the very short time region after the initial disturbance we can
separate the one-particle
and two-particle relaxation. On this time scale the memory in the distribution
functions can be neglected $\bar f_a(\bar t)=f_a(0)$, $\exp\left\{-{t-\bar t\over\tau_g}\right\}=1$ but we will keep the spectral relaxation implicit
in the off-shell $\cos$-function of (\ref{kinetic}).

The eq. (\ref{kinetic}) can be integrated with respect to time and the resulting equation for $f(t)$ represents
the deviation of Wigner's distribution from its initial value, $f_a(t)=
f_a(0)+\delta f_a(t)$, and reads
\begin{eqnarray}\label{short1}
\delta f_a(t)&=&2\sum\limits_b
\int\frac{dpdq}{(2\pi\hbar)^6}V^2_{\rm D}(q)
{1-\cos\left\{{1\over\hbar}t\Delta_E\right\}\over\Delta_E^2}
\left\{f'_a f'_b \bar f_a \bar f_b-f_a
f_b \bar f'_a \bar f'_b\right\}.
\end{eqnarray}
This formula shows how the two-particle and the single-particle concepts
of the transient behavior meet in the kinetic equation. The right hand
side describes how two particles correlate their motion to avoid the
strong interaction regions. Since the process
is very fast, the on-shell contribution to $\delta f_a$, proportional
to $t/\tau$, can be neglected in the assumed time domain and
the $\delta f$ has the pure off-shell character as can be seen from
the off-shell factor $\Delta_E^{-2}\left(1-
\cos\left\{{1\over\hbar}t\Delta_E\right\}\right)$.
The off-shell character of mutual
two-particle correlation is thus reflected in the single particle
Wigner's distribution.

The
very fast formation of the off-shell contribution to Wigner's
distribution has been found in numerical treatments of Green's
functions \cite{D841,K95}. Once formed, the off-shell contributions
change in time with the characteristic time $\tau$, i.e., following
the relaxation (on-shell) processes in the system. Accordingly, the
formation of the off-shell contribution signals that the system has
reached the state the evolution of which can be described by the
nonlocal Boltzmann equation \cite{SLM96}, i.e., the transient time period has been accomplished.

\begin{figure}
\centerline{\parbox[t]{6cm}{
  \psfig{figure=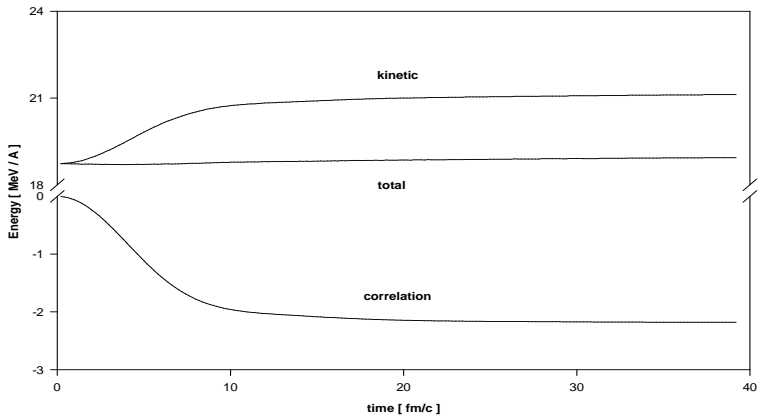,width=10cm,height=6cm,angle=-90}}}
\caption{\label{5}The time evolution of the correlation
and kinetic energy for two
initially counter -flowing streams of
nuclear matter with density and temperature $n_1=n_o/60
\quad
T_1=0.5 $ MeV
and $n_2=n_o/20 \quad T_2=0.1$ MeV moving with relative
velocity of $1
\hbar/fm$ which corresponds to a collision energy of
$21$ MeV/n. In this case the correlation time $\tau_c=9 fm/c$.
The equilibrated temperature is here
(neglecting correlations)  $T=11.6$ MeV and a Gau\ss{} type of potential 
has been used.}
\end{figure}

From Wigner's distribution (\ref{short1}) one can readily evaluate the increase of the
kinetic energy
\begin{eqnarray}
E^{\rm static}_{\rm corr}(t)&=&-\sum_{ab}\int\frac{dkdpdq}{(2\pi\hbar)^9}V^2_{\rm D}(q)
\frac{1-\cos\left\{{1\over\hbar}t\Delta_E\right\}}{\Delta_E}
f'_a f'_b \bar f_a \bar f_b.
\label{energ1}
\end{eqnarray}
This expression holds for general distributions $f_a$.
We choose a model of two
initially counter -
flowing streams of
nuclear matter and plot in figure \ref{5} the time
evolution of
the kinetic energy and the correlation.

This build up of correlations is found to be independent of the form of the
initial distribution. If for example we choose a (equilibrium)
Fermi
distribution as the initial distribution
a build up of correlations will occur as well.
This is due to the fact that the spatial correlations relate in momentum
space to excitations,
resulting in a distribution looking somewhat like a Fermi-distribution
but with a temperature higher than that of the initial uncorrelated
Fermi-distribution.\cite{hsk95}

\section{The formation of correlations}

We start now to calculate the observed increase of kinetic energy analytically. 
As a first example
we shall consider a Yukawa-type potential of the form
$V_{\rm Y}(r)={g_{ab} \over r} {\rm e}^{-\kappa r}$
where in nuclear physics applications $g_{ab}$ is the
coupling and $\kappa$ the effective range of potential
given by the inverse mass of interchanging mesons.
As a second example we shall use
a Gau\ss{}-type potential
$V_{G}(r)=V_0 {\rm e}^{-(r/\eta)^2}$
which has been used in nuclear physics applications\cite{D84,hsk95}
with $\eta=0.57 {\rm fm}^{-3}$ and $V_0=-453$ MeV.

The high temperature limit of the time dependence of the correlation energy 
can be calculated analytically \cite{MK97,MSL97a}. 
The low temperature value is of special interest,
because it leads to a natural definition of the build up of
correlations. One obtains
\beq\label{v3}
{\pa t} E_{\rm corr} &=&
-\frac 1 2 m^4 p_f \langle {V^2 \over \cos{\frac \theta
2}} \rangle \tilde I_f
\label{to1}
\eeq
with standard notation of angular
integrals \cite{SHJ89} and finds the time dependence of the correlation
energy \cite{MK97}
\beq
&&E^{low}_{\rm corr}(t)-E_{\rm corr}^{low}(0)= E_{\rm corr}^{low}
(1 +  \frac 1 3 ({ \epsilon_f+\epsilon_c \over \pi T})^2)^{-1}
\nonumber\\
&\times&
\left \{ 1-\frac 1 x \sin(x)
+\left ({\epsilon_f +\epsilon_c \over \pi T}\right )^2 \left (\frac 1 3 + \left [\frac 1 x \sin(x)\right ]''\right )\right \}
\label{corrt}
\eeq
with $x=2{\epsilon_f+\epsilon_c  \over \hbar} t$ and the equilibrium correlation energy
$E_{\rm corr}^{\rm low}$ for the Yukawa- or Gau\ss{}-
potential reads \cite{MK97}
\beq
E_{\rm corr}&=&\frac 1 6 s_1 s_2 \tilde \epsilon_f \left (T^2+ \frac 1 3 \left ({4 \tilde \epsilon_f \over \pi}\right
)^2\right ) \left ({m \over \hbar^2}\right )^4
%\nonumber\\
%&\times& 
\left \{\matrix{ {g^2\over 2 \kappa^3 \pi^2} \left (
\arctan{\frac {1}{ b_l}}+{b_l \over 1+b_l^2} \right )\quad
Yuk.\cr
{V_0^2 \eta^5 \over 2 \sqrt{2 \pi}} {\rm erf}({p_f \eta \sqrt{2} \over \hbar})\quad \mbox{\it
Gau\ss{}}} \right .\nonumber\\
&&
\label{equil}
\eeq
where $\tilde \epsilon_f={\epsilon_f +\epsilon_c \over 4}$
and $b_l={\hbar \kappa \over 2 p_f}$.
The best choice for
cut-off we found $\epsilon_c \approx \epsilon_f$.
We see that the correlation energy
is built up and oscillates around the equilibrium value damped
with $t^{-1}$ in time. We now define the build up
time $\tau_c$ as the time where the correlation
energy has reached its first maximum. This time is given by
$\tau_c \approx{\hbar \over \epsilon_F}$.

We assert here that the result (\ref{corrt}) is valid
for any binary interaction. We could use Born or $T$-matrix
approximation and the same time dependence
but different $E_{\rm corr}^{low}$ would result.
This correlation time limits the validity of quasiparticle picture which is established at times greater than $\tau_c$ \cite{MSL97a}.
Incidentally, in the early 1950s the criterion $
\hbar/k_BT<\tau$
was supposed to limit the validity of the Landau
Fermi-liquid
theory for metals \cite{P55}. Later it was
shown by Landau that
this criterion is irrelevant and he proposed the
correct
criterion $\tau > \hbar/\epsilon_F$.

\begin{figure}[h]
%\epsfxsize=8cm
%\epsffile{sig1_1.eps}
\centerline{ \parbox[t]{6cm}{
 \psfig{figure=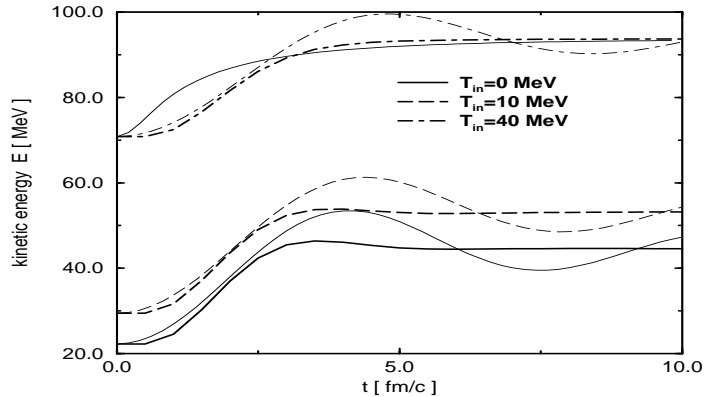,width=10cm,height=6cm,angle=-90}}}
\caption{\label{ill}
The formation of correlation plotted as an increase of the kinetic energy with time for temperatures $1,10,40$ MeV. The
chemical potential is fixed to $37.1$ MeV which corresponds to
densities $0.16, 0.18,0.35 {\rm fm}^{-3}$ . The thick lines show results
from KB calculations while the thin lines
are approximate values via formula (\protect\ref{corrt}).
The equilibrium correlation energy was chosen to be equal to the
KB result. The oscillations are overestimated by the approximate formula. For $T=40$ MeV we plotted also the high temperature approximate value 
\protect\cite{MK97} as thin solid line. The built up of correlations is too fast.}
\end{figure}

In order to illustrate the weak temperature dependence of $\tau_{c}$
we plot in figure \ref{ill} (thick lines)
results from the solution of the Kadanoff
and Baym equations for a fixed chemical potential of $37.1$ MeV
and for three different temperatures. The figure shows
the increase of the kinetic energy (equivalent to the decrease of correlation
energy) with time.
The KB results are compared with those from
approximation (\protect\ref{corrt}). One sees that the agreement is good initially while
correlations are built up. At low temperatures the
oscillations discussed above are obvious in the approximate results
while the KB calculations only show a slight overshoot at the lowest
temperatures.
We believe that the discrepancy is due to the
damping that is neglected in (\ref{short1}).

We can give a Pad{´e} formula for the increase of kinetic energy which interpolates the results of K/B equation within the temperature range up to Fermi energy $\zeta_{\rm K/B}$ and the analytical result (\ref{corrt}) signed $\zeta(t)$
\beq
\zeta_{\rm K/B}(t)&=&{E_{\rm corr}(t) \over E_{\rm corr}(\infty)}=1 - {{1.5225\,\left( 1 - \cos (0.7337\,{x}) +
        \sin (0.6568\,{x}) \right) }\over
    {{{\rm e}^{0.5469\,{x}}}\,{x}}}\label{f}\nonumber\\
\zeta(t)&=&1-{\sin(x) \over x}
\eeq
with $x=4 \epsilon_F t/ \hbar$. The $\zeta_{\rm K/B}$ is a smooth curve through $\zeta(t)$.
As we pointed out above, this result is universal for any binary interaction
approximation. Therefore we can adopt this scaling as a universal
scaling of the one-particle Wigner distribution. Provided we have solved
any local Markovian kinetic equation like BUU, we can
simply multiply the obtained Wigner function by the time
dependent factor (\ref{f}) in order to incorporate the transient
time effect approximately. Since we fixed this factor to the
observable kinetic energy, we expect to have correctly described the
energy variables.

\section{Application to interferometry}

Now we employ the found scaling of short time effects to the
formula (\ref{fp}) and adopt a space-time source
parameterization of
\beq
g(p,r,t)=\rho_0 \Theta(r) \Theta(R_s-r) \Theta(t) p {\rm
e}^{-{p^2 \over 2 m T}} {\rm e}^{-t/\tau} \zeta(t)
\eeq
with the radius $R_s$ and lifetime $\tau$ of the source and the scaling $\zeta(t)$ of (\ref{f}). Without
center of mass momentum $P$ the correlation function becomes
independent of lifetime and collapse to the standard
interferrometry result \cite{JBS86}
\beq
R(0,q)+1={3 \over 4 \pi R_s^3}\int\limits_0^{2 R_s} d^3r
|\varphi(q,r)|^2 \left (1- \frac 3 4 {r\over R_s} +\frac{1}{16}
\left ({r\over R_s}\right )^3\right ).
\eeq
Considering the momentum dependence and assuming initially free
particles $|\varphi(p,r)|^2\propto 1\pm \cos(rq)$ we obtain a
separable representation
\beq
R(P,q)+1=1\pm \Psi(q.R_s) \Pi(\zeta,q.p,\tau)\label{co}
\eeq
with
\beq
\Psi(x)&=&{3\over 2 x^6} \left ((-3+3 x^2-2 x^3+4
x^4)\cos(2x)+(-6 x -6 x^3 -4 x^4)\sin(2x)\right )\nonumber\\&&
\eeq
and
\beq
\Pi(\zeta,q.p/2m,\tau)={\int\limits_0^{\infty}d t dt'{\rm
e}^{-(t+t')/\tau}\zeta(t)\zeta(t') \cos(q.p(t'-t)/2m) \over \left
(\int\limits_0^{\infty} dt {\rm e}^{-t/\tau} \zeta(t)\right )^2}.
\label{pi}
\eeq

\begin{figure}
\centerline{\parbox[t]{5cm}{
  \psfig{figure=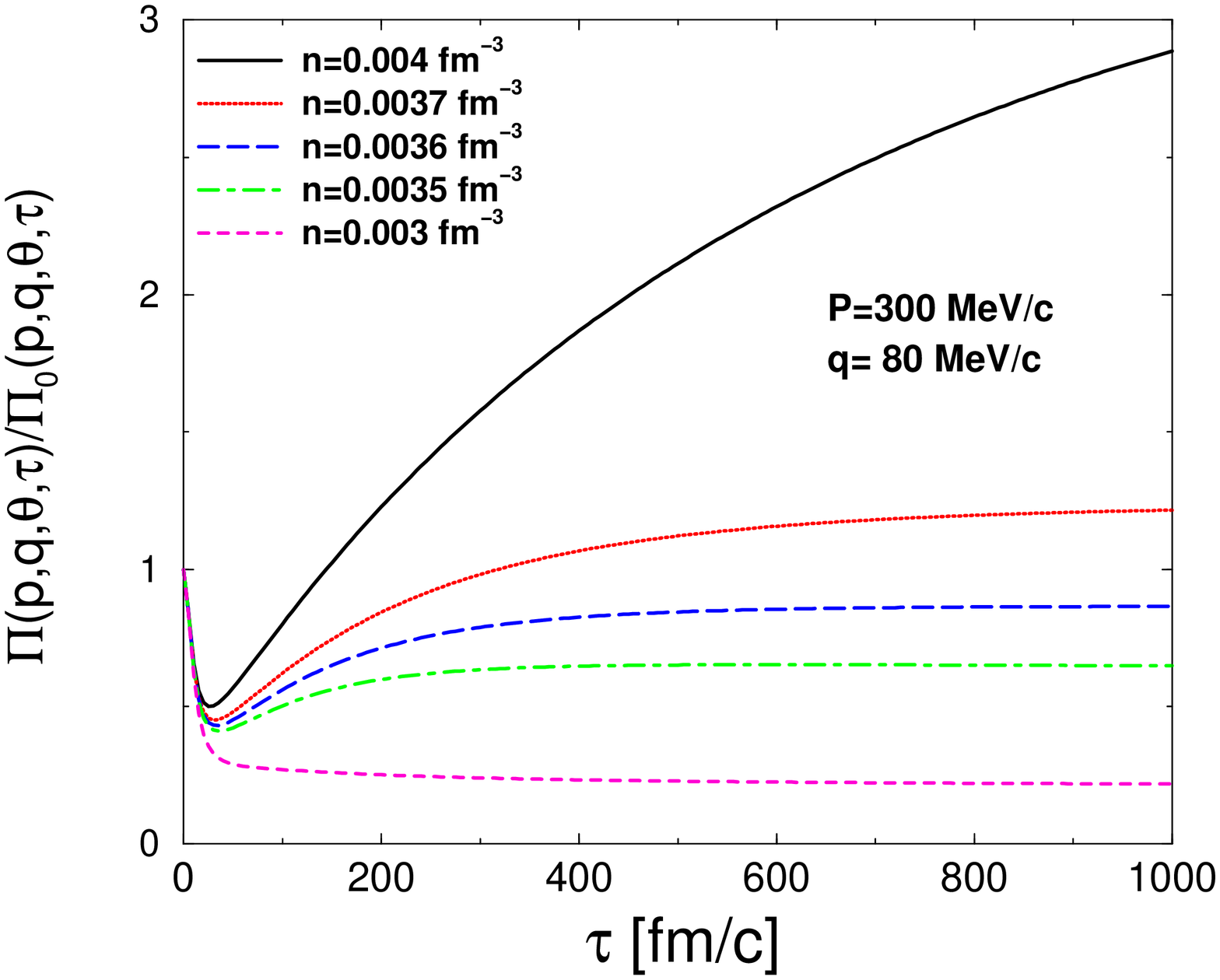,width=6cm,height=5cm,angle=-90}}
\parbox[t]{5cm}{
  \psfig{figure=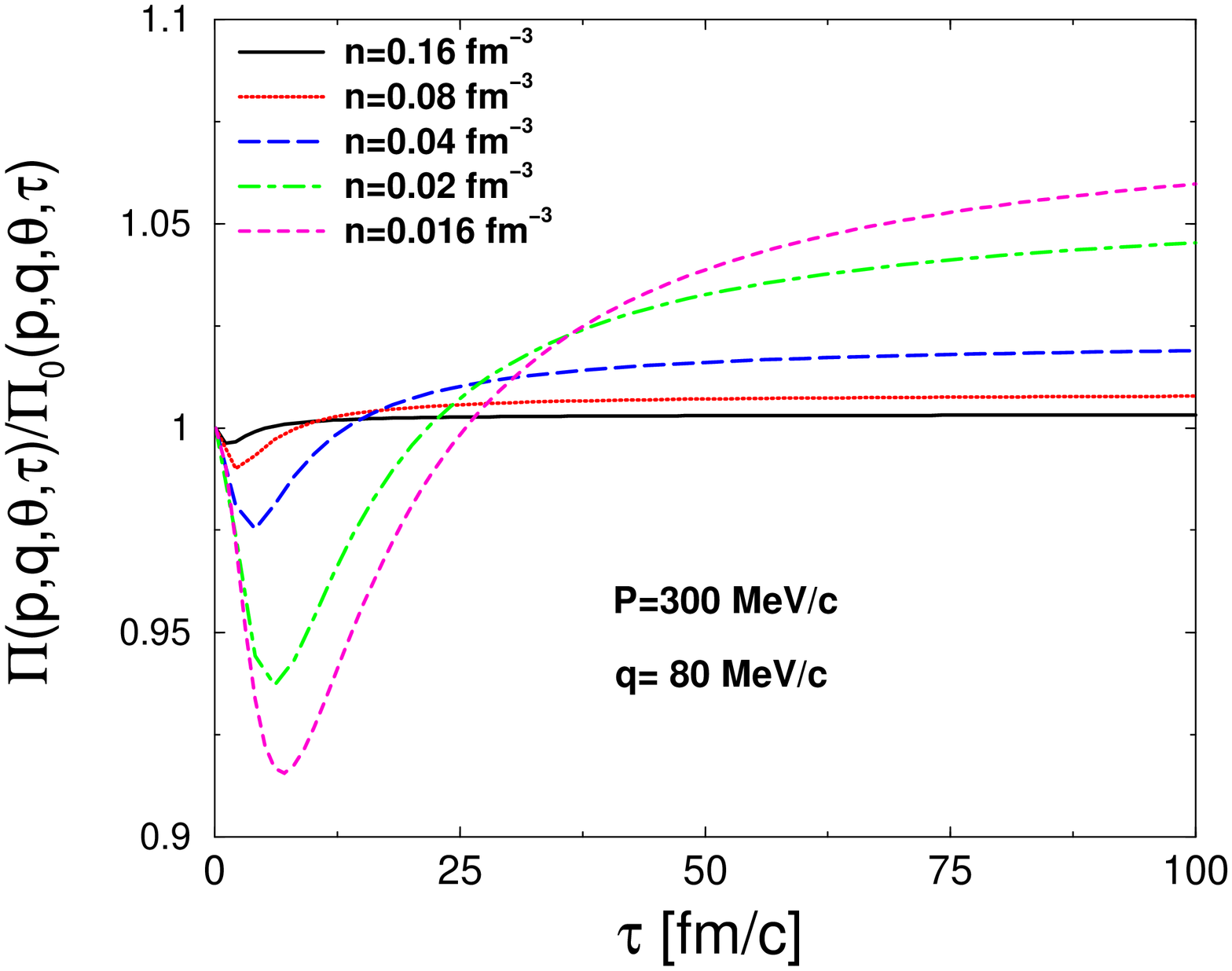,width=6cm,height=5cm,angle=-90}}}
\caption{\label{fac}
The ratio of correlation function $\Pi$ of (\protect\ref{pi})
with short time dynamics to $\Pi_0$ of (\protect\ref{pi0}) without short 
time dynamics versus lifetime of the source.}
\end{figure}

While the first factor $\Psi$ gives the dependence of the
correlation function on the source size, the second factor
contains the effect of finite lifetimes. The result
without transient time effects is that with increasing lifetime
the correlation functions are diminished \cite{BGPG91}. One gets for $\zeta=1$
just 
\beq
\Pi_0(1,q.p,\tau)={1 \over 1+(q.p \tau/2m)^2}.
\label{pi0}
\eeq
We can write the analytical result for (\ref{pi}) with $\zeta(t)$ from (\ref{f})
\beq
\Pi(a,c,\tau)&=&\left (4\,{{\left( {{-2\,a\,\tau }\over {1 + {c^2}\,{{\tau }^2}}} + 
           \arctan (\left( a - c \right) \,\tau ) + 
           \arctan (\left( a + c \right) \,\tau ) \right) }^2} \right .
\nonumber\\
&+& \left .
       {{\left( {{-4\,a\,c\,{{\tau }^2}}\over 
            {1 + {c^2}\,{{\tau }^2}}} - 
          \log (1 + {{\left( a - c \right) }^2}\,{{\tau }^2}) + 
          \log (1 + {{\left( a + c \right) }^2}\,{{\tau }^2})
           \right) }^2}\right )\nonumber\\
&/& 
   (16\,{{\left( a\,\tau  - \arctan (a\,\tau ) \right) }^2})
\eeq
with $c=4 \epsilon_F/\hbar$ and $a=q.p/2m$.
Figure \ref{fac} shows the influence of the transient
time effect on the correlation function. We have plotted
conveniently the ratio of reduced correlation functions $\Pi$ of (\ref{pi})
with short time dynamics to $\Pi_0$ of (\ref{pi0}) without short time dynamics.

We recognize that there is a suppression for smaller lifetimes. Since the
effect of lifetime $\tau$ itself lowers the correlation function
we can state that the extracted lifetimes from experiments are
too high if short time or transient time effects are neglected.
For higher lifetimes this effect is opposite. The extracted lifetimes from experiment should be too low.
Interesting is the density dependence. While first with decreasing density the
lowering/enhancement shifts to higher lifetimes, there is a critical density where we have only suppression. It is found to be around $0.0036 {\rm fm}^{-3}$.

Due to the found universal scaling this effect is important in any
recent simulation of BUU- type since the transient time
effect of formation of correlations and therefore increase of
kinetic energy has not been considered so far. For moleculardynamical methods
this is in principle included but buried in artificial
initial correlations due to the numerical set up. For a comparison
between the kinetic and MD simulation see \cite{MSL97a}.

\section{Summary}

The gradient approximation of the kinetic equation in second order Born
approximation is investigated.
A finite duration approximation of the non Markovian collision integral
is proposed which follows from time dependent Fermi's
Golden Rule and which is in good agreement with the numerical solution of
complete
Kadanoff and Baym equation. 

The build up time of correlations is investigated and it is found that
the low temperature value is universal for any approximation at the
binary collision level. It is shown that the formation time of
correlations is nearly determined by the ratio of $\hbar$ to the transfer
energy which can be considered as an analogue to the uncertainty principle.

A universal scaling is formulated which allows to map the Wigner
function of Markovian simulation to a result including the
transient time effect. As a consequence the influence to
interferrometry methods is discussed. It is found that the
extracted lifetimes from experiment should be too large for lower lifetimes 
and too small for higher values depending on the freeze out-density if transient time effects
are neglected.

\section*{Acknowledgments}
The authors like to thank P. Lipavsk{\'y} and V. {\v S}pi{\v c}ka for interesting discussions and N. Kwong for many useful hints.

\end{document}